\documentstyle[preprint,aps,epsf,floats]{revtex}
\begin{document}

\draft
 
{\tighten
\preprint{\vbox{\hbox{CALT-68-2041}
                \hbox{UCSD/PTH 96-02}
                \hbox{hep-ph/9602262} }}
 
\title{Perturbative corrections to zero recoil inclusive $B$ decay sum rules}
 
\author{Anton Kapustin, Zoltan Ligeti and Mark B. Wise}

\address{California Institute of Technology, Pasadena, CA 91125}

\author{Benjamin Grinstein}

\address{Department of Physics, University of California at San Diego, 
  La Jolla, CA 92093}

\maketitle 

\begin{abstract}%
Comparing the result of inserting a complete set of physical states in a time
ordered product of $b$ decay currents with the operator product expansion gives
a class of zero recoil sum rules.  They sum over physical states with
excitation energies less than $\Delta$, where $\Delta$ is much greater than the
QCD scale and much less than the heavy charm and bottom quark masses.  These
sum rules have been used to derive an upper bound on the zero recoil limit of
the $B\to D^*$ form-factor, and on the matrix element of the kinetic energy
operator between $B$ meson states.  Perturbative corrections to the sum rules
of order $\alpha_s(\Delta)\,\Delta^2/m_{c,b}^2$ have previously been computed. 
We calculate the corrections of order $\alpha_s(\Delta)$ and
$\alpha_s^2(\Delta)\,\beta_0$ keeping all orders in $\Delta/m_{c,b}$, 
and show that these perturbative QCD corrections suppressed by powers of
$\Delta/m_{c,b}$ significantly weaken the upper bound on the zero recoil $B\to
D^*$ form-factor, and also on the kinetic energy operator's matrix element.

\end{abstract}

}

\newpage

Over the last six years, dramatic progress has been achieved in our
understanding of exclusive and inclusive $B$ decays.  For exclusive decays this
resulted from applying heavy quark symmetry \cite{HQS} to relate $B$ decay
form-factors and obtain their normalization at zero recoil.  For example, the
form-factors that occur in $B\to D\,e\,\bar\nu_e$ and $B\to D^*\,e\,\bar\nu_e$
semileptonic decays are related by heavy quark symmetry to a single universal
function of $v\cdot v'$ ($v$ is the four-velocity of the $B$, and $v'$ is that
of the recoiling $D^{(*)}$), and furthermore, this function is normalized to
unity at zero recoil \cite{HQS,NuWe,VoSi,Luke}.  

Progress in the theory of inclusive $B$ decays has come from applying the
operator product expansion and heavy quark effective theory \cite{eft} to
perform a $1/m_b$ expansion of the time ordered product of $b$ decay currents
\cite{CGG}.  It was found that at leading order in this expansion, the
inclusive semileptonic $B$ decay rate is equal to the perturbative $b$ quark
decay rate.  There are no nonperturbative corrections at order $1/m_b$, and the
corrections of order $1/m_b^2$ are characterized by only two matrix elements
(we use the standard relativistic normalization for the $B$ meson states)
\begin{equation}
\lambda_1 = {1\over2m_B}\, \langle B(v) \,|\, \bar h_v^{(b)}\, (iD)^2\,
  h_v^{(b)} \,|\, B(v)\rangle \,,
\end{equation}
and 
\begin{equation}
\lambda_2 = {1\over6m_B}\, \langle B(v) \,|\, \bar h_v^{(b)}\, {g\over2}\,
  \sigma_{\mu\nu}\, G^{\mu\nu}\, h_v^{(b)} \,|\, B(v)\rangle \,,
\end{equation}
where $h_v^{(b)}$ is the $b$ quark field in the heavy quark effective theory
\cite{incl,MaWi,BKSV}.  The matrix element $\lambda_2$ is scale dependent
\cite{l2run}, and it is determined from the measured $B^*-B$ mass splitting, 
$\lambda_2(m_b)\simeq0.12\,{\rm GeV}^2$.

Sum rules have been derived that relate exclusive decay form-factors to the
matrix elements $\lambda_{1,2}$ \cite{BSUV}.  The zero recoil sum rules 
follow from analysis of the time ordered product
\begin{equation}
T_{\mu\nu} = {i\over2m_B}\, \int{\rm d}^4x\, e^{-iq\cdot x}\,
  \langle B \,|\, T\{J_\mu^\dagger(x),J_\nu(0)\} | B\, \rangle \,,
\end{equation}
where $J_\nu$ is a $b\to c$ axial or vector current, the $B$ states
are at rest, $\vec q=0$ and $q^0=m_b-m_c-\epsilon$.
Viewed as a function of complex $\epsilon$, $T_{\mu\nu}$ has two cuts
along the real $\epsilon$-axis.  One, for $\epsilon\gtrsim0$, corresponds
to physical states with a charm quark and the other, for 
$\epsilon\lesssim-2m_c$, corresponds to physical intermediate states with two
$b$ quarks and a $\bar c$ quark.  The first cut arises from inserting the 
states between the two currents in the product $J^\dagger J$, and the second 
cut arises from inserting the states between the currents in the other time
ordering $JJ^\dagger$.  So we arrive at
\begin{eqnarray}
T_{\mu\nu}(\epsilon) &=& {1\over2m_B}\, \sum_X\, (2\pi)^3\,\delta^3(\vec p_X)\,
  {\langle B| J_\mu^\dagger |X\rangle \langle X| J_\nu |B\rangle \over
  (m_b-m_c)-(m_B-m_X)-\epsilon-i\,0} \nonumber\\
&-& {1\over2m_B}\, \sum_X\, (2\pi)^3\,\delta^3(\vec p_X)\,
  {\langle B| J_\nu |X\rangle \langle X| J_\mu^\dagger |B\rangle \over
  (m_b-m_c)+(m_B-m_X)-\epsilon+i\,0} \,.
\end{eqnarray}
The sum over $X$ includes the usual phase space factors, {\it i.e.}, 
${\rm d}^3p/2E$ for each particle in the state $X$.

\begin{figure}[t]
\centerline{\epsfysize=9truecm \epsfbox{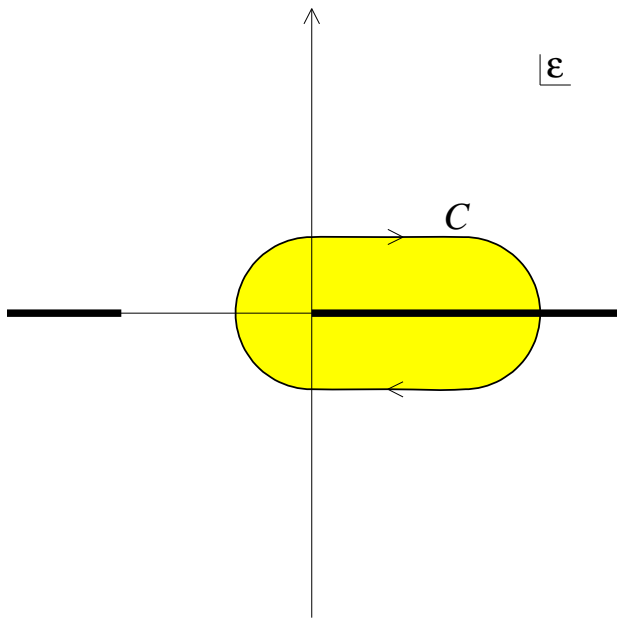}}
\caption[1]{The integration contour $C$ in the complex $\epsilon$ plane.
The cuts extend to ${\rm Re}\,\epsilon\to\pm\infty$. }
\end{figure}

Consider integration of the product of a weight function $W_\Delta(\epsilon)$
with $T_{\mu\nu}(\epsilon)$ along the contour $C$ shown in Fig.~1.  Assuming
$W$ is analytic in the shaded region enclosed by this contour and averaging
over $\mu=\nu=1,2,3$, we get 
\begin{equation}\label{weighted}
{1\over2\pi i}\, \int_C\, {\rm d}\epsilon\, W_\Delta(\epsilon)\, 
  {T_{ii}(\epsilon)\over3}
= \sum_X\, W_\Delta[(m_X-m_c)-(m_B-m_b)]\, (2\pi)^3\,\delta^3(\vec p_X)\,
  {\Big|\langle X| J_i |B\rangle\Big|^2\over3\cdot2m_B}\, .
\end{equation}
The maximum $X$ mass on the right-hand side of eq.~(\ref{weighted}) is
determined by where the contour $C$ pinches the real axis.  For convenience
this mass is chosen to be less than $2m_b+m_c$ to prevent the occurrence of
states $X$ with $b$, $\bar b$, and $c$ quarks.  We take the maximum $X$ mass to
be $2m_B$.  Hereafter it is understood that sums over $X$ only go over states
up to mass $2m_B$.

We require that: ($i$) the weight function $W_\Delta$ be positive semidefinite
along the cut so that every term in the sum over $X$ on the right-hand side of
eq.~(\ref{weighted}) is non-negative; ($ii$) $W_\Delta(0)=1$; ($iii$)
$W_\Delta$ be flat near $\epsilon=0$ ({\it i.e.}, at least
${\rm d}W_\Delta(\epsilon)/{\rm d}\epsilon\Big|_{\epsilon=0}=0$); ($iv$) and
that it falls off rapidly to zero for $\epsilon>\Delta$.  We want to take
$\Delta\ll m_{c,b}$.  Then states $X$ other than the $D^*$ give a contribution 
to the right-hand side of eq.~(\ref{weighted}) that is suppressed by 
$(1/m_{c,b})^2$.  However, in our numerical results we consider $\Delta$ as 
large as $2\,$GeV.  Although our analysis holds for any weight function that 
satisfies these four properties, for explicit calculations we use
\begin{equation}\label{weightfn}
W_\Delta^{(n)}(\epsilon) = {\Delta^{2n}\over\epsilon^{2n}+\Delta^{2n}} \,,
\end{equation}
with $n=2,3,\ldots$ (for $n=1$ the integral over $\epsilon$ is dominated by
contributions from states with mass of order $m_B$).  These weight functions
have poles at $\epsilon=\root{2n}\of{-1}\,\Delta$, therefore, as long as $n$ 
is not too large and $\Delta$ is much larger than the QCD scale, 
$\Lambda_{\rm QCD}$, the contour in Fig.~1 is far from the cut until $\epsilon$
is near $2m_B$.  Then we should be able to calculate the integral in
eq.~(\ref{weighted}) using the operator product expansion to evaluate the time
ordered product.  

The choice of the set of weight functions in eq.~(\ref{weightfn}) is motivated
by the fact that for values of $n$ of order unity all poles of $W_\Delta^{(n)}$
lie at a distance of order $\Delta$ away from the physical cut.  In this case
the integral along the contour $C$ can be computed only assuming local duality
\cite{PQW} at the scale $2m_B$.  The dependence of our results on this
assumption is extremely weak, because for $\Delta\ll m_B$ the weight function
is very small where the contour $C$ touches the cut.  As $n\to\infty$,
$W_\Delta^{(n)}$ approaches $\theta(\Delta-\epsilon)$ for positive $\epsilon$,
which corresponds to summing over all hadronic resonances up to excitation
energy $\Delta$ with equal weight.  Then the poles of $W_\Delta^{(n)}$ approach
the cut, and the contour $C$ is forced to lie within distance of order
$\Delta/n$ from the cut at $\epsilon=\Delta$.  In this case the evaluation of
the integral along the contour $C$ relies also on local duality at the scale
$\Delta$.\footnote{In fact, for any sequence of functions analytic in some
neighbourhood of the positive real axis that converges to
$\theta(\Delta-\epsilon)$, some singularity will approach $\epsilon=\Delta$. 
Thus, the pinching of the contour is inevitable if one uses a weight function
that varies rapidly.}

Neglecting perturbative QCD corrections and nonperturbative effects 
corresponding to operators of dimension greater than five, the operator 
product expansion gives \cite{BKSV}
\begin{equation}\label{opeAA}
\frac13\, T_{ii}^{AA} = - \frac1\epsilon + {(\lambda_1+3\lambda_2)(m_b-3m_c)
  \over 6m_b^2\,\epsilon\,(2m_c+\epsilon)}
  - {4\lambda_2 m_b-(\lambda_1+3\lambda_2)(m_b-m_c-\epsilon)\over
  m_b\,\epsilon^2\,(2m_c+\epsilon)} \,,
\end{equation}
when $J_\mu=A_\mu=\bar c\,\gamma_\mu\gamma_5\,b$, and
\begin{equation}\label{opeVV}
\frac13\, T_{ii}^{VV} = - \frac1{2m_c+\epsilon} +
  {(\lambda_1+3\lambda_2)(m_b+3m_c)\over 6m_b^2\,\epsilon\,(2m_c+\epsilon)}
  - {4\lambda_2 m_b-(\lambda_1+3\lambda_2)(m_b-m_c-\epsilon)\over
  m_b\,\epsilon\,(2m_c+\epsilon)^2} \,,
\end{equation}
when $J_\mu=V_\mu=\bar c\,\gamma_\mu\,b$.  
Performing the contour integration yields
\begin{mathletters}\label{nonpert}
\begin{eqnarray}
\frac1{6m_B}\, && \sum_X\, W_\Delta[(m_X-m_c)-(m_B-m_b)]\, (2\pi)^3\,
  \delta^3(\vec p_X)\, \Big|\langle X| A_i |B\rangle\Big|^2 \nonumber\\*
&& = 1 - {\lambda_2\over m_c^2} + \bigg({\lambda_1+3\lambda_2\over4}\bigg)
  \bigg(\frac1{m_c^2}+\frac1{m_b^2}+\frac2{3m_c m_b}\bigg) \,, \label{npA}\\
\frac1{6m_B}\, && \sum_X\, W_\Delta[(m_X-m_c)-(m_B-m_b)]\, (2\pi)^3\,
  \delta^3(\vec p_X)\, \Big|\langle X| V_i |B\rangle\Big|^2 \nonumber\\*
&& = {\lambda_2\over m_c^2} - \bigg({\lambda_1+3\lambda_2\over4}\bigg)
  \bigg(\frac1{m_c^2}+\frac1{m_b^2}-\frac2{3m_c m_b}\bigg) \,. \label{npV}
\end{eqnarray}
\end{mathletters}%
These equations hold for any $W_\Delta$ that satisfies the four properties
mentioned above.  Higher order terms in the operator product expansion
for $T_{ii}$ give contributions with more factors of $1/\epsilon$ on the 
right-hand sides of eqs.~(\ref{opeAA}) and (\ref{opeVV}).  Therefore, if the 
weight function has nonvanishing $m$'th derivative at $\epsilon=0$,
there are corrections to the right-hand side of eq.~(\ref{npA}) of order
\begin{equation}
\bigg({\Lambda_{\rm QCD}\over m_{c,b}}\bigg)
  \bigg({\Lambda_{\rm QCD}\over\Delta}\bigg)^{\!m} =
  \bigg({\Lambda_{\rm QCD}\over m_{c,b}}\bigg)^{\!2}\, \bigg[
  \bigg({m_{c,b}\over\Lambda_{\rm QCD}}\bigg) 
  \bigg({\Lambda_{\rm QCD}\over\Delta}\bigg)^{\!m} \bigg] \,.
\end{equation}
We require that $\Delta$ be large enough compared with the QCD scale
$\Lambda_{\rm QCD}$, so that such terms are smaller than those we kept in
eq.~(\ref{npA}).  For $m>1$ $\Delta$ can still be smaller than $m_{c,b}$. 
Higher order terms in the operator product expansion of $T_{ii}^{VV}$ give
corrections to the right-hand side of eq.~(\ref{npV}) of order 
$(\Lambda_{\rm QCD}/m_{c,b})^2\,(\Lambda_{\rm QCD}/\Delta)^{m-1}$.  This is why
we imposed condition ($iii$).  For the weight function
$W^{(n)}_\Delta(\epsilon)$ in eq.~(\ref{weightfn}) the first nonvanishing
derivative is at $m=2n$.  

We have considered the nonperturbative corrections to the sum rules
(\ref{nonpert}) characterized by $\lambda_1$ and $\lambda_2$.  There are also
perturbative corrections suppressed by powers of the strong coupling.  These
are most easily calculated not in the operator product expansion, but by
directly considering the sum over states in (\ref{nonpert}) and replacing the
hadronic states by quark and gluon states.  The perturbative corrections are of
two types.  There are corrections of order $\alpha_s(m_{c,b})$ not suppressed
by powers of $\Delta/m_{c,b}$.  These arise, at the parton level, from the
final state $X=c$ and change the term $1$ on the right-hand side of (\ref{npA})
to $\eta_A^2$, where $\eta_A$ is the usual factor that relates the axial
current in the full theory of QCD to the axial current in the heavy quark 
effective theory (at zero recoil).  $\eta_A$ has been calculated to order 
$\alpha_s$ \cite{VoSi}, and terms of order $\alpha_s^2\,\beta_0$, where
$\beta_0=(11-\frac23\,n_f)$, are also known \cite{eta2,else}.  Explicitly,
\begin{equation}\label{etaA}
\eta_A = 1 - {\alpha_s\over\pi}\, \bigg({m_b+m_c\over m_b-m_c}\ln{m_c\over m_b}
  + \frac83 \bigg) - {\alpha_s^2\over\pi^2}\, \beta_0\, \frac5{24}\,
  \bigg({m_b+m_c\over m_b-m_c}\ln{m_c\over m_b} + \frac{44}{15} \bigg) \,,
\end{equation}
where $\alpha_s$ is the $\overline{\rm MS}$ coupling evaluated at the scale
$\sqrt{m_bm_c}$.

There is another class of perturbative QCD corrections coming from final states
$X$ that contain a charm quark plus additional partons, {\it e.g.}, $c\,g$,
$c\,\bar q\,q$, {\it etc}.  They give a contribution to the right-hand side of
equations (\ref{nonpert}) that is of order
$[\alpha_s(\Delta)+\ldots]\,F(\Delta)$, where the ellipses denote terms of
higher order in the strong coupling constant $\alpha_s$, and for small
$\Delta$, $F(\Delta)\sim\Delta^2/m_{c,b}^2$.  We have evaluated the strong
coupling constant at the scale $\Delta$, because that characterizes the typical
hadronic mass in the sum over $X$.  Note that although these corrections are
suppressed by powers of $\Delta/m_{c,b}$ they can be as important as the other
perturbative corrections we considered, since the strong coupling constant is
evaluated at a lower scale $\Delta$.  The value for these corrections depends
on the precise form of the weight function, and we use the ones given in
eq.~(\ref{weightfn}).  Such perturbative corrections were calculated at order
$\alpha_s(\Delta)\,\Delta^2/m_{c,b}^2$ in the limit when the weight function
approaches the step-function $\theta(\Delta-\epsilon)$ (corresponding to
$W_\Delta^{(\infty)}$) \cite{BSUV,KMY}.  As we have already pointed out,
the use of such a weight function relies on local duality at the scale
$\Delta$, so the corrections are expected to be less than those stemming from
$W_\Delta^{(n)}$ with small $n$ relying on local duality only at the scale
$2m_B$.  We calculated for $n\geq2$ the terms of order $\alpha_s(\Delta)$
coming from the Feynman diagrams in Fig.~2, and the order
$\alpha_s^2(\Delta)\,\beta_0$ terms arising from the diagrams shown in Fig.~3. 
Then eqs.~(\ref{npA}) and (\ref{npV}) become

\begin{figure}[tb]
\centerline{\epsfysize=3truecm \epsfbox{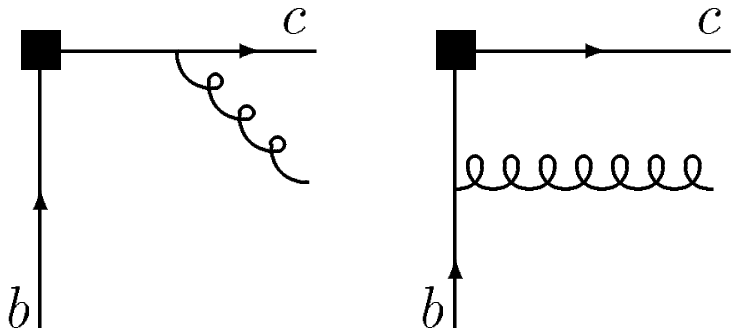}}
\caption[2]{Feynman diagrams that contribute to the order $\alpha_s(\Delta)$ 
corrections to the sum rules.  
The black square indicates insertion of the $b\to c$ axial or vector current.}
\end{figure}

\begin{figure}[tb]
\centerline{\epsfysize=3truecm \epsfbox{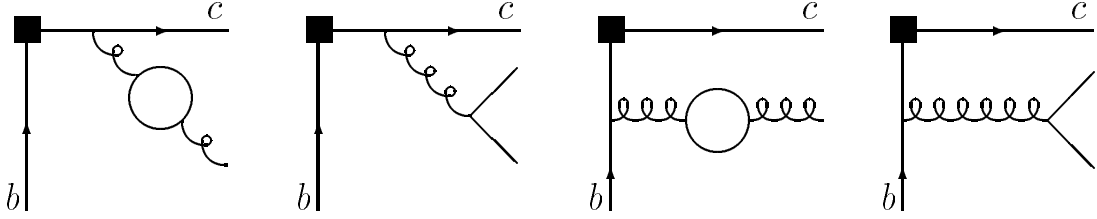}}
\caption[3]{Feynman diagrams that determine the order 
$\alpha_s^2(\Delta)\,\beta_0$ corrections to the sum rules. }
\end{figure}

\begin{mathletters}\label{sumrules}
\begin{eqnarray}
\frac1{6m_B}\, && \sum_X W_\Delta^{(n)}[(m_X-m_c)-(m_B-m_b)]\, (2\pi)^3\,
  \delta^3(\vec p_X)\, \Big|\langle X| A_i |B\rangle\Big|^2  \label{sumrA} \\*
=&& \eta_A^2 - {\lambda_2\over m_c^2} + 
  \bigg({\lambda_1+3\lambda_2\over4}\bigg)
  \bigg(\frac1{m_c^2}+\frac1{m_b^2}+\frac2{3m_c m_b}\bigg) 
  + {\alpha_s(\Delta)\over\pi}\, X_{AA}^{(n)}(\Delta) + 
  {\alpha_s^2(\Delta)\over\pi^2}\,\beta_0\, Y_{AA}^{(n)}(\Delta) \,,\nonumber\\
\frac1{6m_B}\, && \sum_X W_\Delta^{(n)}[(m_X-m_c)-(m_B-m_b)]\, (2\pi)^3\,
  \delta^3(\vec p_X)\, \Big|\langle X| V_i |B\rangle\Big|^2  \label{sumrV} \\*
=&& {\lambda_2\over m_c^2} - \bigg({\lambda_1+3\lambda_2\over4}\bigg)
  \bigg(\frac1{m_c^2}+\frac1{m_b^2}-\frac2{3m_c m_b}\bigg)  
  + {\alpha_s(\Delta)\over\pi}\, X_{VV}^{(n)}(\Delta) + 
  {\alpha_s^2(\Delta)\over\pi^2}\,\beta_0\, Y_{VV}^{(n)}(\Delta) \,. \nonumber
\end{eqnarray}
\end{mathletters}%
On the right-hand sides of eqs.~(\ref{sumrules}) terms suppressed by more than
two powers of $\Lambda_{\rm QCD}/m_{c,b}$ or $\alpha_s$ have been neglected. 
We have also neglected in (\ref{sumrA}) terms suppressed by 
$(\Lambda_{\rm QCD}/m_{c,b})\,(\Lambda_{\rm QCD}/\Delta)^{2n}$ and in
(\ref{sumrV}) terms suppressed by 
$(\Lambda_{\rm QCD}/m_{c,b})^2\,(\Lambda_{\rm QCD}/\Delta)^{2n-1}$. 
Perturbative corrections to the terms proportional to $\lambda_{1,2}$ are also
neglected, and we evaluate $\lambda_2$ in eqs.~(\ref{sumrules}) at the scale
$m_b$ (a calculation of QCD corrections to its coefficient would resolve this
scale ambiguity).  For $\Delta\ll m_{c,b}$ the functions $X^{(n)}$ and 
$Y^{(n)}$ are given by 
\begin{eqnarray}
X_{AA}^{(n)}(\Delta) &=& \Delta^2\, {A^{(n)}\over3}\,
  \bigg(\frac1{m_c^2}+\frac1{m_b^2}+\frac2{3m_c m_b}\bigg) \,, \nonumber\\*
X_{VV}^{(n)}(\Delta) &=& \Delta^2\, {A^{(n)}\over3}\,
  \bigg(\frac1{m_c^2}+\frac1{m_b^2}-\frac2{3m_c m_b}\bigg) \,, \nonumber\\*
Y_{AA}^{(n)}(\Delta) &=& \Delta^2\, {B^{(n)}\over6}\,
  \bigg(\frac1{m_c^2}+\frac1{m_b^2}+\frac2{3m_c m_b}\bigg) 
  + \Delta^2\, {A^{(n)}\over15}\, 
  \bigg(\frac1{m_c^2}+\frac1{m_b^2}+\frac{4}{3m_c m_b}\bigg) \,, \nonumber\\*
Y_{VV}^{(n)}(\Delta) &=& \Delta^2\, {B^{(n)}\over6}\,
  \bigg(\frac1{m_c^2}+\frac1{m_b^2}-\frac2{3m_c m_b}\bigg) \,,
\end{eqnarray}
where the coefficients $A^{(n)}$ and $B^{(n)}$ are 
($n\geq2$)
\begin{equation}
A^{(n)} = {\pi\over n\sin(\pi/n)} \,, \qquad
  B^{(n)} = A^{(n)}\, \bigg( {\pi\over2n\tan(\pi/n)} 
  + \frac53 - \ln2 \bigg) \,.
\end{equation}

For $\Delta$ near $1\,$GeV higher powers of $\Delta/m_{c,b}$ are important.
The analytic expressions for $X^{(\infty)}_{AA}$ and $X^{(\infty)}_{VV}$ are
(for $\Delta<2m_b$)
\begin{eqnarray}
X_{AA}^{(\infty)} &=& {\Delta\,(\Delta+2m_c)\,
  [2(\Delta+m_c)^2-2m_b^2-(m_b+m_c)^2] \over18m_b^2\,(\Delta+m_c)^2} + 
  {3m_b^2+2m_bm_c-m_c^2\over9m_b^2}\, \ln{\Delta+m_c\over m_c} \,,\nonumber\\*
X_{VV}^{(\infty)} &=& {\Delta\,(\Delta+2m_c)\,
  [2(\Delta+m_c)^2-2m_b^2-(m_b-m_c)^2] \over18m_b^2\,(\Delta+m_c)^2} + 
  {3m_b^2-2m_bm_c-m_c^2\over9m_b^2}\, \ln{\Delta+m_c\over m_c} \nonumber\,. \\*
\end{eqnarray}
($X_{AA}^{(\infty)}$ was also calculated in Ref.~\cite{KMY}.  Our result seems
to disagree with theirs.) In Figures 4a and 4b we plot $X^{(\infty)}$ and
$Y^{(\infty)}$ versus $\Delta$ using the values $m_b=4.8\,$GeV and
$m_c=1.4\,$GeV.  The thick solid lines are $X$ and the thick dashed lines are
$Y$, while the thin lines are the corresponding functions at order
$\Delta^2/m_{c,b}^2$.  Note that expanding in $\Delta/m_{c,b}$ is not a good
approximation unless $\Delta\lesssim400\,$MeV.

\begin{figure}[t]
\centerline{\epsfysize=8truecm \epsfbox{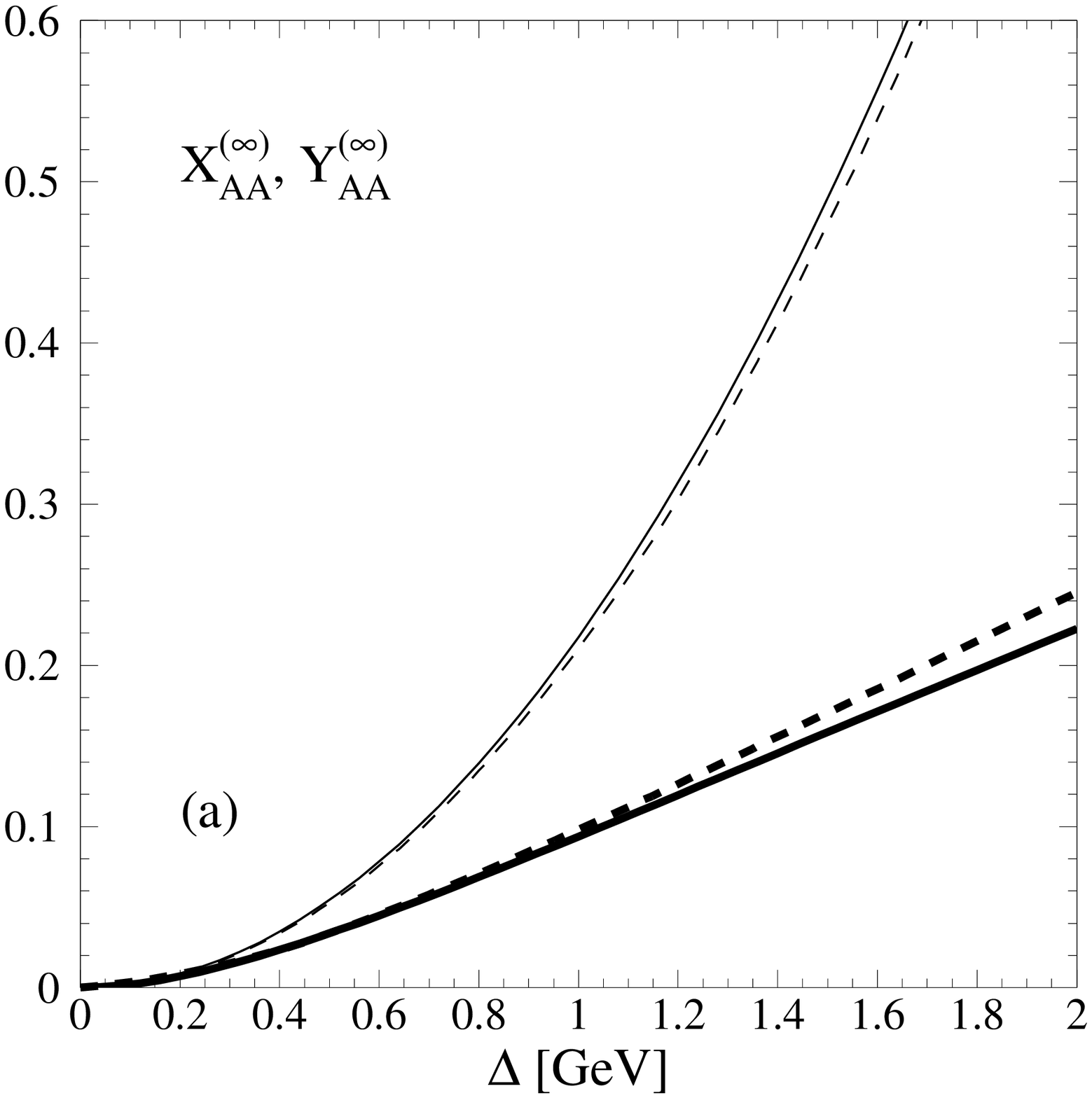} 
  \epsfysize=8truecm \epsfbox{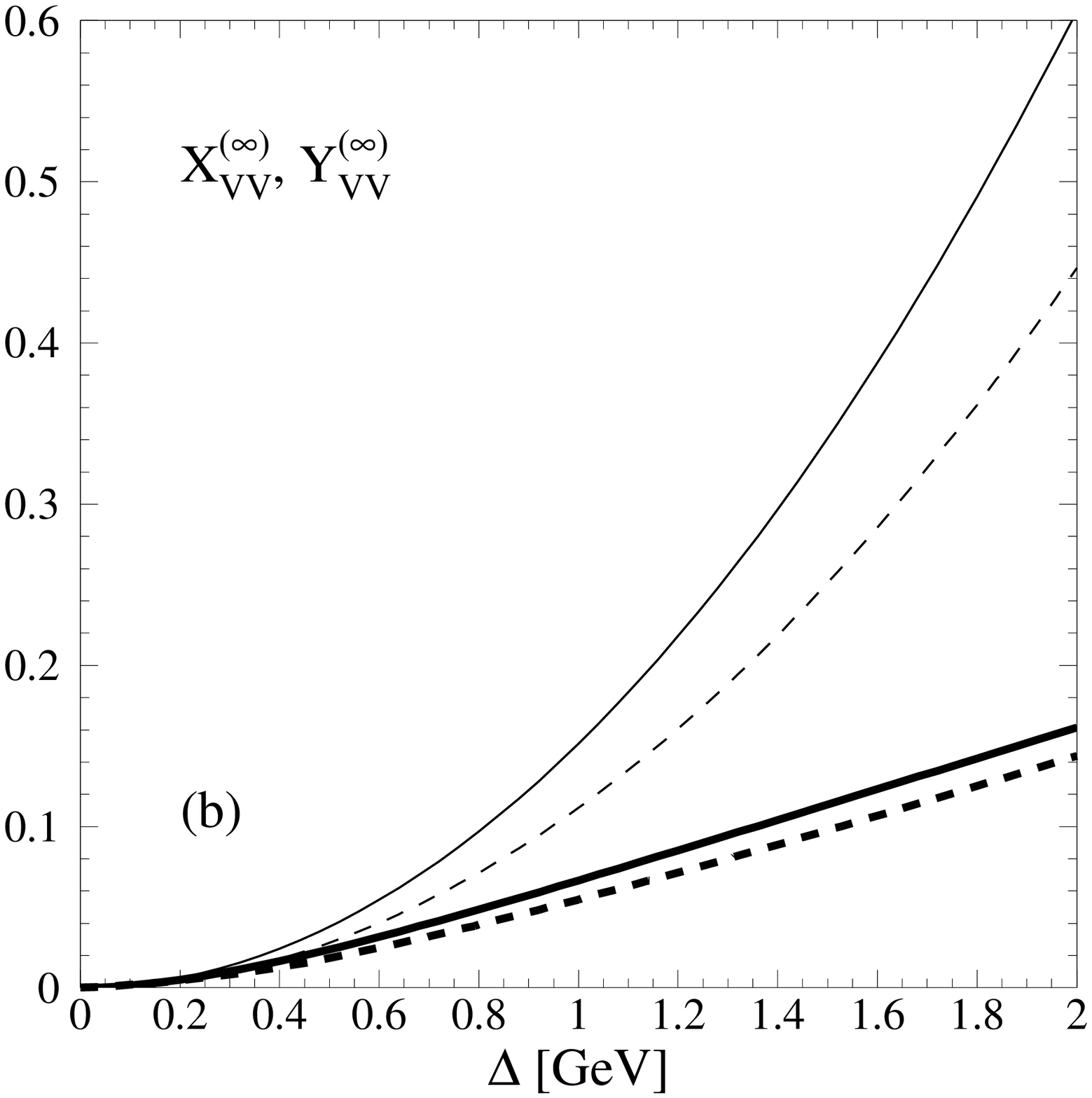}}
\caption[4]{$X^{(\infty)}(\Delta)$ and $Y^{(\infty)}(\Delta)$ for the 
a) axial, and b) vector coefficients.  Thick solid lines are $X$ while 
thick dashed lines are $Y$.  The thin solid and dashed lines are $X$
and $Y$ to order $\Delta^2/m_{c,b}^2$. }
\end{figure}

The evaluation of the order $\alpha_s^2\,\beta_0$ corrections is made
relatively simple by the relation between the $n_f$ dependent part of the order
$\alpha_s^2$ contribution and the order $\alpha_s$ contribution with a finite
gluon mass \cite{SmVo}.  Such a relation holds in the so-called $V$-scheme, but
throughout this paper we present all results in the usual $\overline{\rm MS}$
scheme.  Knowledge of the order $\alpha_s^2\,\beta_0$ corrections allows us to
obtain the BLM scale \cite{BLM} that results from absorbing vacuum polarization
effects into the running coupling constant.  It is generally believed that this
choice of scale yields a reasonable perturbative expansion.  This is also the
reason for using $\alpha_s(\Delta)$ in the sum rules in eqs.~(\ref{sumrules}). 
Had we chosen some very different scale $\mu$, the coefficients $Y^{(n)}$ would
contain large logarithms of $\mu^2/\Delta^2$.  Using $m_b=4.8\,$GeV,
$m_c=1.4\,$GeV, and $n\to\infty$ we obtain $\mu_{\rm BLM}^{AA}\simeq0.12\Delta$
and $\mu_{\rm BLM}^{VV}\simeq0.17\Delta$ for the BLM scales for the axial and
vector current sum rules, respectively.  If we demand 
$\alpha_s(\mu_{\rm BLM})<1$, then $\Delta$ needs to be above about $3-4\,$GeV,
which would completely eliminate the restrictive power of the sum rules. 
Another possibility, which we adopt, is just to use our results as estimates of
the $\alpha_s^2$ corrections, but to retain $\Delta\simeq1\,$GeV.  Then we find
that the order $\alpha_s^2(\Delta)$ corrections to the sum rules are comparable
to the order $\alpha_s(\Delta)$ terms, and we can only hope that terms of
higher order in $\alpha_s(\Delta)$ are not similarly important.  

The parameters $\lambda_1$ and $\lambda_2$ also occur in the inclusive 
differential $B$ decay rate, which can be expressed in terms of the $B$
and $D$ meson masses and the parameters $\lambda_1$, $\lambda_2$ and
$\bar\Lambda$, where
\begin{equation}\label{masses}
m_B = m_b + \bar\Lambda - {\lambda_1+3\lambda_2\over2m_b}\,, \qquad
m_D = m_c + \bar\Lambda - {\lambda_1+3\lambda_2\over2m_c}\,.
\end{equation}
The pole mass is not a physical quantity, and the perturbative expression for
the $\overline{\rm MS}$ mass $\overline{m_b}(m_b)$ in terms of the pole mass
$m_b$ is not Borel summable, giving rise to what is sometimes called a
``renormalon ambiguity" in the pole mass \cite{renorm}.  However, when the
differential semileptonic decay rate is expressed in terms of the hadron masses
and $\bar\Lambda$, the perturbative QCD corrections to the decay rate are also
not Borel summable.  If $\bar\Lambda$ (or equivalently the $b$ quark pole mass)
extracted from the differential semileptonic decay rate is used to get the
$\overline{\rm MS}$ mass these ambiguities cancel, so one can arrive at a
meaningful prediction for the $\overline{\rm MS}$ $b$ quark mass.  It is fine
to introduce unphysical quantities like $\bar\Lambda$ as long as one works
consistently to a given order of QCD perturbation theory and the expansion in
inverse powers of the heavy quark masses.  Since the final results one
considers always involve relations between physically measurable quantities,
any ``renormalon ambiguities" arising from the bad behavior of the QCD
perturbation series at large orders will cancel out \cite{rencan,NeSa}.  As the
left-hand sides of the sum rules in eqs.~(\ref{sumrules}) are physical
quantities, the right-hand sides, when calculated to all orders in $\alpha_s$,
should be free of renormalon ambiguities.  We checked that the order
$\Lambda_{\rm QCD}^2/m_{c,b}^2$ renormalon ambiguity in the quantity $\eta_A^2$
\cite{NeSa} cancels against that in the perturbative corrections suppressed by
$\Delta^2/m_{c,b}^2$ (such a cancellation was conjectured in \cite{conj}).

Eqs.~(\ref{npA}) and (\ref{npV}) have been used to bound the $B\to D^*$ zero 
recoil form-factor \cite{SUV,BSUV}.  The sum rule (\ref{sumrA}) implies 
a bound on the zero recoil $B\to D^*$ matrix element of the axial current 
$F_{B\to D^*}^2$, defined by
$\langle D^*|A_i|B\rangle=2\sqrt{m_{D^*}m_B}\,F_{B\to D^*}\,\varepsilon_i$,
that reads
\begin{eqnarray}\label{finalA}
F_{B\to D^*}^2 &\leq& \eta_A^2 - {\lambda_2\over m_c^2} 
  + \bigg({\lambda_1+3\lambda_2\over4}\bigg)  
  \bigg(\frac1{m_c^2}+\frac1{m_b^2}+\frac2{3m_c m_b}\bigg) \nonumber\\*
&+& {\alpha_s(\Delta)\over\pi}\, X_{AA}^{(n)}(\Delta) + 
  {\alpha_s^2(\Delta)\over\pi^2}\,\beta_0\, Y_{AA}^{(n)}(\Delta) \,.
\end{eqnarray}
Here we used that the contributions of states $X$ of higher mass than the 
$D^*$ to the left-hand side of (\ref{sumrA}) are positive, and neglected the 
very small deviation of $W_\Delta^{(n)}[(m_{D^*}-m_c)-(m_B-m_b)]$ from unity
implied by eq.~(\ref{weightfn}), eq.~(\ref{masses}), and the relation
$m_{D^*}-m_D=2\lambda_2/m_c$.  The positivity of the sum over states $X$ in
eq.~(\ref{sumrV}) implies that
\begin{equation}
0 \leq {\lambda_2\over m_c^2} - \bigg({\lambda_1+3\lambda_2\over4}\bigg)
  \bigg(\frac1{m_c^2}+\frac1{m_b^2}-\frac2{3m_c m_b}\bigg) 
  + {\alpha_s(\Delta)\over\pi}\, X_{VV}^{(n)}(\Delta) + 
  {\alpha_s^2(\Delta)\over\pi^2}\,\beta_0\, Y_{VV}^{(n)}(\Delta) \,.
\end{equation}
This inequality gives a constraint on the heavy quark effective theory matrix 
element $\lambda_1$, which is strongest when one takes the 
$m_c\gg m_b\gg\Delta$ limit, giving
\begin{equation}\label{finalV}
\lambda_1 \leq -3\lambda_2 +
  {\alpha_s(\Delta)\over\pi}\,\Delta^2\, {4A^{(n)}\over3} + 
  {\alpha_s^2(\Delta)\over\pi^2}\,\beta_0\,\Delta^2\, {2B^{(n)}\over3} \,.
\end{equation}

\begin{table}[b] 
  \begin{tabular}{cc|ddd} 
&  &  $n=2$ &  $n=3$  &  $n=\infty$  \\  \hline 
$\lambda_1$  &  to order $\alpha_s$  &  $-0$.$06\,$GeV$^2$  &
  $-0$.$13\,$GeV$^2$  &  $-0$.$17\,$GeV$^2$  \\ 
&  to order $\alpha_s^2\,\beta_0$  &  $0$.$13\,$GeV$^2$  & $0$.$06\,$GeV$^2$  
  &  $0$.$01\,$GeV$^2$  \\ \hline
$F_{B\to D^*}$  &  to order $\alpha_s$  &  0.96  &  0.96  &  0.95  \\ 
&  to order $\alpha_s^2\,\beta_0$  &  0.99  &  0.99  &  0.98 
\end{tabular} \vskip6pt
\caption[]{Upper limits on $\lambda_1$ and $F_{B\to D^*}$ that can be obtained
from the sum rules in eqs.~(\ref{finalA}) and (\ref{finalV}) with
$\Delta=1\,$GeV.  $n$ labels the weight function $W_\Delta^{(n)}$. } 
\end{table}

Neglecting the perturbative corrections suppressed by powers of
$\alpha_s(\Delta)$, eq.~(\ref{finalV}) yields $\lambda_1\leq-0.36\,{\rm
GeV}^2$, which in turn implies using eq.~(\ref{finalA}) that $F_{B\to
D^*}\leq0.93$ \cite{SUV}.  (With $m_b=4.8\,$GeV and $m_c=1.4\,$GeV we find
$\eta_A=0.96$ following from eq.~(\ref{etaA}).)  To indicate the importance of
the perturbative corrections proportional to $\alpha_s(\Delta)$ and
$\alpha_s^2(\Delta)\,\beta_0$, we give the bounds that result when they are
included for $n=2$, $n=3$, and $n\to\infty$ in Table~I.  The effects of these
corrections are smaller if we choose $\Delta$ small (corresponding to
suppressing the contribution of higher excited states) or if we choose $n$
large (using local duality at the scale $\Delta$).  Note that while it is
plausible that $n$ can be chosen arbitrarily large as local duality is expected
to hold at scales much above $\Lambda_{\rm QCD}$, the relation
$\Delta\gg\Lambda_{\rm QCD}$ must be maintained and so $\Delta$ cannot be
chosen to be less than about $1\,$GeV.  Using  $n_f=3$, $\Delta=1\,$GeV and
$\alpha_s(1\,{\rm GeV})=0.45$ we obtain the bounds given in Table~I.  The large
magnitude of the second order corrections to the sum rules indicates that the
series of perturbative corrections might be under control only for $\Delta$
significantly above $1\,$GeV.  Such a value for $\Delta$ would greatly weaken
the restrictive power of the sum rules.  Similar comments and conclusions apply
to two analogous sum rules derived for $B^*\to D^{(*)}$ transitions in
Ref.~\cite{MNsr}.

In conclusion, we investigated perturbative corrections to the zero recoil
inclusive $B$ decay sum rules derived in Ref.~\cite{BSUV}.  We calculated the
corrections suppressed by powers of $\Delta/m_{c,b}$ at order
$\alpha_s(\Delta)$ and order $\alpha_s^2(\Delta)\,\beta_0$ corresponding to a
set of possible weight functions that determine the contributions of excited
hadronic intermediate states.  These corrections significantly weaken the
constraints stemming from the sum rules.  It is widely believed that
$\lambda_1<0$ (although in our opinion it has not been proven in QCD for
$\lambda_1$ defined by the $\overline{\rm MS}$ subtraction scheme), and we are
not aware of any claim that $F_{B\to D^*}$ is significantly above 1.  Due to
the size of the $\Delta$-dependent terms in the sum rules, it is hard to deduce
any useful model independent bounds.  An upper bound below 1 on the zero recoil
$B\to D^*$ form-factor barely survives these perturbative corrections, and a
limit on $\lambda_1$ that restricts it to negative values does not.  However,
it is important to remember that the results in Table~I rely on the
applicability of QCD perturbation theory at a scale $\Delta=1\,$GeV, and
furthermore are very sensitive to the value of $\alpha_s$ at this scale.  In
the future $\lambda_1$ may be determined from experimental data on inclusive
$B$ decays \cite{pheno}, and then a bound on $F_{B\to D^*}$ that does not rely
on eq.~(\ref{finalV}) can be derived from eq.~(\ref{finalA}).  

In light of our discussion, we see no reason to think that the original
estimates of $F_{B\to D^*}$, based on model calculations, the structure of
terms arising at order $1/m_{c,b}^2$ \cite{estim1}, and on chiral perturbation
theory \cite{estim2}, badly underestimated the $1/m_{c,b}^2$ corrections.  Our
results show  that the zero recoil sum rules do not demand a larger deviation
of $F_{B\to D^*}$ from $\eta_A$, even if the $D^*$ does not saturate the sum
over states $X$.  We cannot prove at this point that such a deviation does not
occur.  However, in the absence of any such indication, it is most natural to
think that $F_{B\to D^*}\simeq\eta_A=0.96$ holds to an accuracy of about the
canonical size of the $1/m_{c,b}^2$ corrections, that is within $(500\,{\rm
MeV}/2m_c)^2\simeq3\%$.

\acknowledgements
We thank Aneesh Manohar for useful discussions.  
This work was supported in part by the U.S.\ Dept.\ of Energy under Grant no.\
DE-FG03-92-ER~40701 and contract DOE-FG03-90ER40546.  The research of 
B.\ G.\ was also supported in part by the Alfred P.\ Sloan Foundation.
A.\ K.\ was supported in part by the Schlumberger Foundation.

\newpage
{\tighten
 
} 

\end{document}